\documentclass[twocolumn,prb,showpacs,superscriptaddress,groupedaddress]{revtex4}
\usepackage{amsmath}
\usepackage{dcolumn}

\usepackage[pdfborder={0 0 0},colorlinks=true,linkcolor=blue,citecolor=blue]{hyperref}

\usepackage{graphicx}
\usepackage{rotating}
\usepackage{amssymb}
\usepackage{mathptmx}

\newcommand{\sign}{\mathrm{sign}}

\begin{document}

\title{Impurity-induced smearing of the spin resonance peak in Fe-based superconductors}



\author{Yu.N. Togushova}
\affiliation{Siberian Federal University, Svobodny Prospect 79, Krasnoyarsk 660041, Russia}

\author{V.A. Shestakov}
\affiliation{Siberian Federal University, Svobodny Prospect 79, Krasnoyarsk 660041, Russia}

\author{M.M. Korshunov}
\email{mkor@iph.krasn.ru}
\affiliation{Siberian Federal University, Svobodny Prospect 79, Krasnoyarsk 660041, Russia}
\affiliation{L.V. Kirensky Institute of Physics, Krasnoyarsk 660036, Russia}

\date{\today}


\begin{abstract}
The spin resonance peak in the iron-based superconductors is observed in inelastic neutron scattering experiments and agrees well with predicted results for the extended $s$-wave ($s_\pm$) gap symmetry. On the basis of four-band and three-orbital tight binding models we study the effect of nonmagnetic disorder on the resonance peak. Spin susceptibility is calculated in the random phase approximation with the renormalization of the quasiparticle self-energy due to the impurity scattering in the static Born approximation. We find that the spin resonance becomes broader with the increase of disorder and its energy shifts to higher frequencies. For the same amount of disorder the spin response in the $s_\pm$ state is still distinct from that of the $s_{++}$ state.
\end{abstract}

\maketitle

\section{Introduction}

Discovery of Fe-based superconductors (FeBS) in 2008 with the maximal $T_c$ of 55K gave rise to the debates on the origin of the superconducting state. FeBS can be broadly divided into the two classes, pnictides and chalcogenides~\cite{Reviews}. 
Since conductivity is provided by the FeAs layer, the discussion of physics in terms of quasi two-dimensional system in most cases gives reasonable results~\cite{ROPPreview2011}.
Fe $d$-orbitals form the Fermi surface (FS) that excluding the cases of extreme hole and electron dopings consists of two hole sheets around the $\Gamma=(0,0)$ point and two electron sheets around the $M=(\pi,\pi)$ point in the 2-Fe Brillouin zone (BZ). In the 1-Fe BZ, latter corresponds to the electron sheets around the $(\pi,0)$ and $(0,\pi)$ points. Nesting between these two groups of sheets is the driving force for the spin-density wave (SDW) long-range magnetism in the undoped FeBS. Upon doping the SDW state is destroyed but the residual scattering with the wave vector $\mathbf{Q}$ connecting hole and electron pockets naturally leads to the enhanced antiferromagnetic fluctuations. $\mathbf{Q}$ is equal to $(\pi,\pi)$ in the 2-Fe BZ and to $(\pi,0)$ or $(0,\pi)$ in the 1-Fe BZ.

Different mechanisms of Cooper pairs formation result in distinct superconducting gap symmetry and structure~\cite{ROPPreview2011}. In particular, the RPA-SF (random-phase approximation spin fluctuation) approach gives the extended $s$-wave gap that changes sign between hole and electron Fermi surface sheets ($s_{\pm}$ state) as the main instability for the wide range of doping concentrations~\cite{ROPPreview2011,Graser,KorshunovUFN}.
On the other hand, orbital fluctuations promote the order parameter to have the sign-preserving $s_{++}$ symmetry~\cite{Kontani}. Thus, probing the gap structure can help in elucidating the underlying mechanism. In this respect, inelastic neutron scattering is a powerful tool since the measured dynamical spin susceptibility $\chi(\mathbf{q},\omega)$ in the superconducting state carries information about the gap structure.

For the local interactions (Hubbard and Hund's exchange), $\chi$ can be obtained in the RPA from the bare electron-hole bubble $\chi_0(\mathbf{q},\omega)$ by summing up a series of ladder diagrams to give
\begin{eqnarray}
\chi(\mathbf{q},\omega) = \left[I - U_s \chi_0(\mathbf{q},\omega)\right]^{-1} \chi_0(\mathbf{q},\omega),
\label{eq:chi_s_sol}
\end{eqnarray}
where $U_s$ and $I$ are interaction and unit matrices in orbital or band space, and all other quantities are matrices as well. Scattering between nearly nested hole and electron Fermi surfaces in FeBS produce a peak in the normal state magnetic susceptibility at or near $\mathbf{q} = \mathbf{Q}$. For the uniform $s$-wave gap, $\sign \Delta_\mathbf{k} = \sign \Delta_{\mathbf{k}+\mathbf{Q}}$ and there is no resonance peak. For the $s_\pm$ order parameter as well as for an extended non-uniform $s$-wave symmetry, $\mathbf{Q}$ connects Fermi sheets with the different signs of gaps. This fulfills the resonance condition for the interband susceptibility, and the spin resonance peak is formed at a frequency $\omega_s$ below $\Omega_c = \min \left(|\Delta_\mathbf{k}| + |\Delta_{\mathbf{k}+\mathbf{q}}| \right)$. The existence of the spin resonance in FeBS was predicted theoretically \cite{Korshunov2008,Maier} and subsequently discovered experimentally with many reports of well-defined spin resonances in all systems, see~\cite{ROPPreview2011}.

Since there are always some amount of disorder even in the crystals of a very good quality, it is necessary to study the evolution of the spin response with increasing amount of disorder. Here we do this within two models for the band structure -- one is the simple four-band model in the 2-Fe BZ~\cite{Korshunov2008} and the other one is the three-orbital model in the 1-Fe BZ~\cite{Korshunov3orb} with the spin-orbit coupling~\cite{Korshunov3orbSO}. The effect of disorder on the spin susceptibility is incorporated via the static Born approximation for the quasiparticle self-energy due to the impurity scattering.

\section{Models and approximations}

We study the spin response in the superconducting state of FeBS within the tight-binding models for the two-dimensional iron layer. Some basic information can be gained from the four-band model of Ref.~\cite{Korshunov2008}, that is able to reproduce the FS obtained via band structure calculations. It has the following single-electron Hamiltonian
\begin{eqnarray}
H_0 = - \sum\limits_{\mathbf{k},\alpha ,\sigma } {{\epsilon^i} n_{\mathbf{k} i \sigma}} - \sum\limits_{\mathbf{k}, i, \sigma}  t_{\mathbf{k}}^{i} d_{\mathbf{k} i \sigma}^\dag d_{\mathbf{k} i \sigma},
\label{eq:H04band}
\end{eqnarray}
where $d_{\mathbf{k} i \sigma}$ is the annihilation operator of the $d$-electron with momentum $\mathbf{k}$, spin $\sigma$, and band index $i = \left\{ \alpha_1, \alpha_2, \beta_1, \beta_2 \right\}$, $\epsilon^i$ are the on-site single-electron energies, $t_{\mathbf{k}}^{\alpha_{1},\alpha_{2}} = t^{\alpha_{1},\alpha_{2}}_1 \left(\cos k_x+\cos k_y \right)+ t^{\alpha_{1},\alpha_{2}} _2 \cos k_x \cos k_y$ is the electronic dispersion that yields hole pockets centered around the $\Gamma$ point, and $t_{\mathbf{k}}^{\beta_{1},\beta_{2}} = t^{\beta_{1},\beta_{2}}_1 \left(\cos k_x+\cos k_y \right)+ t^{\beta_{1},\beta_{2}}_2 \cos \frac {k_x}{2} \cos \frac{k_y}{2}$ is the dispersion that results in the electron pockets around the $M$ point. Using the abbreviation $(\epsilon^i, t_1^i, t_2^i)$ we choose the parameters $(-0.60,0.30,0.24)$ and $(-0.40,0.20,0.24)$ for the $\alpha_1$ and $\alpha_2$ bands, respectively, and $(1.70,1.14,0.74)$  and $(1.70,1.14,-0.64)$ for the $\beta_1$ and $\beta_2$ bands, correspondingly (all values are in eV).

The matrix elements of the bare spin susceptibility in the multiband system has the form:
%
\begin{eqnarray}
\chi_0^{ij}(\mathbf{q},\mathrm{i} \Omega)&=&-\frac{T}{2} \sum_{\mathbf{k}, \omega_n} \left[ G^i(\mathbf{k} + \mathbf{q}, \mathrm{i} \omega_n + \mathrm{i} \Omega) G^j(\mathbf{k} , \mathrm{i} \omega_n) \right. \nonumber \\
&+& \left. F^i(\mathbf{k} + \mathbf{q}, \mathrm{i} \omega_n + \mathrm{i} \Omega) F^j(\mathbf{k} , {\rm i} \omega_n)\right],
\label{eq:bare_chi}
\end{eqnarray}
where $\Omega$ and $\omega_n$ are Matsubara frequencies, $G^i$ and $F^i$ are the normal and anomalous (superconducting) Green's functions, respectively. Physical spin susceptibility $\chi(\mathbf{q},\mathrm{i}\Omega) = \sum_{i,j}\chi^{i,j}(\mathbf{q},\mathrm{i}\Omega)$ obtained by calculating matrix elements $\chi^{i,j}(\mathbf{q},\mathrm{i}\Omega)$ via equation~(\ref{eq:chi_s_sol}) with the interaction matrix $U_s^{i,j} = \tilde{U} \delta_{i,j} + \tilde{J}/2 (1-\delta_{i,j})$.
We assume here the effective Hubbard interaction parameters to be $\tilde{J}=0.2 \tilde{U}$ and $\tilde{U} \sim t^{\beta_1}_1$ in order to stay in the paramagnetic phase~\cite{Korshunov2008}. We consider the magnetic susceptibility in the superconducting state assuming the $s_\pm$ state with $\Delta_\mathbf{k} = \frac{\Delta_0}{2}\left( \cos k_x + \cos k_y \right)$, where $\Delta_0$ was chosen to be $5$meV.

\subsection{Three-orbital model}

The model described above lack for the orbital content of the bands. Now we introduce the additional level of complexity by considering the three-orbital model in the 1-Fe BZ~\cite{Korshunov3orb}. By introducing the spin-orbit (SO) interaction to it~\cite{Korshunov3orbSO}, it is possible to explain the observed anisotropy of the spin resonance peak in Ni-doped Ba-122~\cite{Lipscombe2010}. In particular, $\chi_{+-}$ and  $2\chi_{zz}$ components of the spin susceptibility are different thus breaking the spin-rotational invariance $\left<S_+ S_-\right> = 2\left<S_z S_z\right>$. This model comes from the three $t_{2g}$ $d$-orbitals. The $xz$ and $yz$ components are hybridized and form two electron-like FS pockets around $(\pi,0)$ and $(0,\pi)$ points, and one hole-like pocket around $\Gamma=(0,0)$ point. The $xy$ orbital is considered to be decoupled from them and form an outer hole pocket around $\Gamma$ point.
Latter differs from some popular orbital models for FeBS~\cite{ROPPreview2011,Graser}. However, according to ARPES data~\cite{Brouet2012,Kordyuk2012} and the DFT calculations for highly doped systems~\cite{Backes2014} and undoped 122, 1111, and 111 materials~\cite{Nekrasov2008}, $xy$ orbital contribution to the Fermi surface near $\Gamma$ point is quite large in the 2-Fe Brillouin zone. This situation is simulated by introducing the $xy$ hole pocket near $\Gamma$ point in the three-orbital model.
The Hamiltonian is given by $H = H_0+H_{SO}$, where $H_0=\sum\limits_{\mathbf{k},\sigma,l,m} \varepsilon_\mathbf{k}^{l m} c_{\mathbf{k} l \sigma}^\dag c_{\mathbf{k} m \sigma}$ is one-electron part with $c_{\mathbf{k} m \sigma}$ being the annihilation operator of a particle with momentum $\mathbf{k}$, spin $\sigma$ and orbital index $m$. Keeping in mind the similarity of $H_0$ to the Sr$_2$RuO$_4$ case, for simplicity we consider only the $L_z$-component of the SO interaction, which affects $xz$ and $yz$ bands only~\cite{Eremin2002}.
The matrix of the full Hamiltonian $H$ has the form
\begin{equation}\label{epsks_z}
 \hat\varepsilon_{\mathbf{k} \sigma} = \left(
                      \begin{array}{ccc}
                        \varepsilon_{1\mathbf{k}} & 0 & 0 \\
                        0 & \varepsilon_{2\mathbf{k}} & \varepsilon_{4\mathbf{k}} + \mathrm{i} \frac{\lambda}{2} \mathrm{sign}{\sigma} \\
                        0 & \varepsilon_{4\mathbf{k}} - \mathrm{i} \frac{\lambda}{2} \mathrm{sign}{\sigma} & \varepsilon_{3\mathbf{k}} \\
                      \end{array}
                     \right),
\end{equation}
where
%
%
%
\begin{eqnarray}\label{eps3orb}
 \varepsilon_{1\mathbf{k}} &=& \epsilon_{xy} - \mu + 2 t_{xy} (\cos{k_x}+\cos{k_y}) + 4 t_{xy}' \cos{k_x}\cos{k_y}, \nonumber\\
 \varepsilon_{2\mathbf{k}} &=& \epsilon_{yz} - \mu + 2 t_x \cos{k_x} + 2 t_y \cos{k_y} + 4 t' \cos{k_x}\cos{k_y} \nonumber\\
    &+& 2 t'' (\cos{2 k_x}+\cos{2 k_y}), \nonumber\\
 \varepsilon_{3\mathbf{k}} &=& \epsilon_{xz} - \mu + 2 t_y \cos{k_x} + 2 t_x \cos{k_y} + 4 t' \cos{k_x}\cos{k_y} \nonumber\\
    &+& 2 t'' (\cos{2 k_x}+\cos{2 k_y}), \nonumber\\
 \varepsilon_{4\mathbf{k}} &=& 4 t_{xzyz} \sin{k_x/2}\sin{k_y/2}. \nonumber
\end{eqnarray}
To reproduce the topology of the FS in pnictides, we choose the following parameters (in eV): $\mu=0$, $\epsilon_{xy}=-0.70$, $\epsilon_{yz}=-0.34$, $\epsilon_{xz}=-0.34$, $t_{xy}=0.18$, $t_{xy}'=0.06$, $t_x=0.26$, $t_y=-0.22$, $t'=0.2$, $t''=-0.07$, $t_{xzyz}=0.38$.
%
As in the case of Sr$_2$RuO$_4$, eigenvalues of $\hat\varepsilon_{\mathbf{k} \sigma}$ do not depend on the spin $\sigma$, therefore, spin-up and spin-down states are still degenerate in spite of the SO interaction.

Components of the physical spin susceptibility $\chi_{+-,zz}(\mathbf{q},\mathrm{i}\Omega) = \frac{1}{2} \sum_{l,m} \chi^{ll,mm}_{+-,zz}(\mathbf{q},\mathrm{i}\Omega)$ are calculated using Eq.~(\ref{eq:chi_s_sol}) with the interaction matrix $U_s$ from Ref.~\cite{Graser}. We choose the following values for the interaction parameters: spin-orbit coupling constant $\lambda=100$meV, intraorbital Hubbard $U=0.9$eV, Hund's $J=0.1$eV, interorbital $U'=U-2J$, and pair-hopping term $J'=J$. In the superconducting state we assume either the $s_{++}$ state with $\Delta_\mathbf{k} = \Delta_0$ or the $s_\pm$ state with $\Delta_\mathbf{k} = \Delta_0 \cos k_x \cos k_y$, where $\Delta_0 = 20$meV.

\subsection{Impurity scattering}

As were shown recently~\cite{Efremov2011,Stanev2012,KorshunovPRB2014},
the multiband superconductors may demonstrate behavior much more complicated than originally expected from the Abrikosov-Gor'kov theory~\cite{AG}. In particular, $s_\pm \to s_{++}$ transition may take place for the sizeable intraband attraction in the two-band $s_\pm$ model with the nonmagnetic impurities~\cite{Efremov2011}. Discussion of such effects are well beyond the scope of the present study since it requires a self-consistent solution of the frequency and gap equations within the strong-coupling $T$-matrix approximation. Here we use a simple static Born approximation for the quasiparticle self-energy to see the basic effects of nonmagnetic disorder on the spin resonance. That is, the multiple scattering on the same impurity results in the following self-energy: $\Sigma(\mathbf{k}) \approx -\frac{\mathrm{i}}{2\tau_\mathbf{k}}$ with $\tau_\mathbf{k}$ being the quasiparticle lifetime (see, e.g. the so-called first Born approximation in Ref.~\cite{BruusFlensberg}). Calculating the exact momentum dependence of the quasiparticle lifetime is again the separate complicated task that would require realistic multiorbital models with proper orbital-to-bands contribution similar to what was done for the calculation of the transport coefficients in Ref.~\cite{Kemper2011}. This is again beyond the scope of the present work, so, neglecting the momentum dependence of $\tau_\mathbf{k}$, we set $\Sigma(\mathbf{k},\mathrm{i}\Omega) = -\mathrm{i}\Gamma$, where we treat the impurity scattering rate $\Gamma$ as a parameter.

\section{Results and discussion}

\begin{figure}[ht]
\begin{center}
\includegraphics[width=1.0\linewidth]{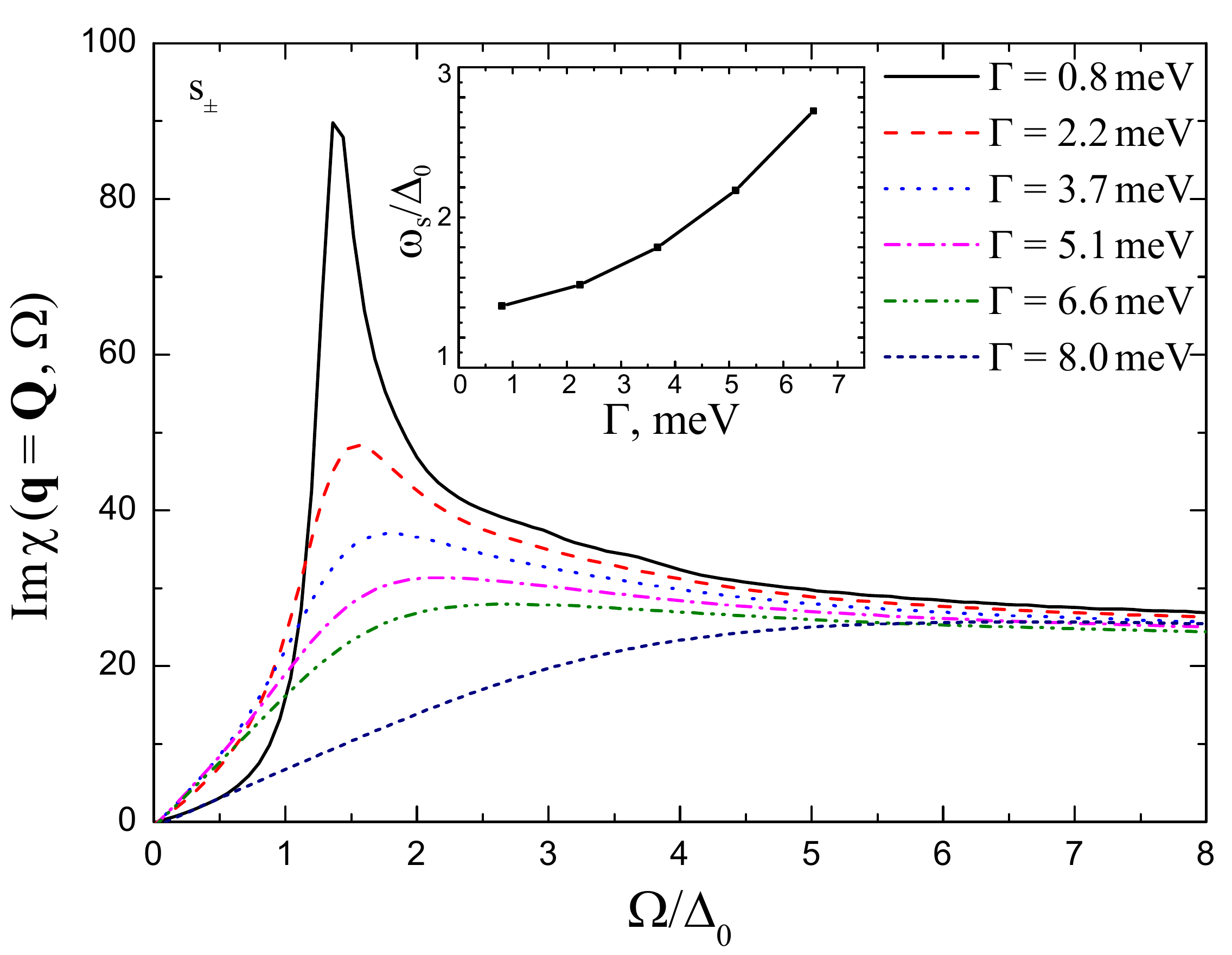}
\end{center}
\caption{Calculated $\mathrm{Im}\chi(\mathbf{Q},\Omega)$ with $\mathbf{Q}=(\pi,\pi)$ in the 2-Fe BZ for the four-band model in the $s_\pm$ state (main panel) and the spin resonance frequency $\omega_s$ determined as the maximum of $\mathrm{Im}\chi(\mathbf{Q},\Omega)$ (inset) for different values of the impurity scattering rate $\Gamma$. The spin resonance below $\Omega = 2\Delta_0$ becomes much broader with increasing $\Gamma$ and effectively disappears for $\Gamma > \Delta_0$.
\label{fig:ImChi4band}}
\end{figure}

First, we consider the spin response in the four-band model. The result of the analytical continuation ($\mathrm{i}\Omega_n \to \Omega + \mathrm{i}\delta$ with $\delta \to 0+$) is show in Fig.~\ref{fig:ImChi4band} for the set of impurity scattering rates $\Gamma$. In the case of small $\Gamma$, the spin resonance peak is clearly seen below the energy of $2\Delta_0$. With increasing $\Gamma$ it becomes broader and almost vanishes once $\Gamma$ becomes comparable to $\Delta_0$. We can trace the energy of the spin resonance $\omega_s$ as a function of $\Gamma$. Value of $\omega_s$ is determined as the maximum of $\mathrm{Im}\chi(\mathbf{Q},\Omega)$. The result is shown in Fig.~\ref{fig:ImChi4band}. Clearly, $\omega_s$ shifts to higher frequencies with increasing disorder.

These findings are in good agreement with the results of Ref.~\cite{MaitiResonance2011} where the band model was simpler then used here but the vertex corrections in the particle-hole bubble due to the impurity scattering were included.
In particular, for the same reduction of the resonance peak height we see similar broadening of the peak and small changes in the resonance frequency.
Such agreement imply that the vertex corrections do not play a crucial role in the low-energy spin response while they are known to be important for the proper calculation of the transport coefficients.
On the other hand, compared to Ref.~\cite{MaitiResonance2011}, we go to larger values of the scattering rate and observe a nonlinear increase of the resonance frequency.

\begin{figure}[ht]
\begin{center}
\includegraphics[width=1.0\linewidth]{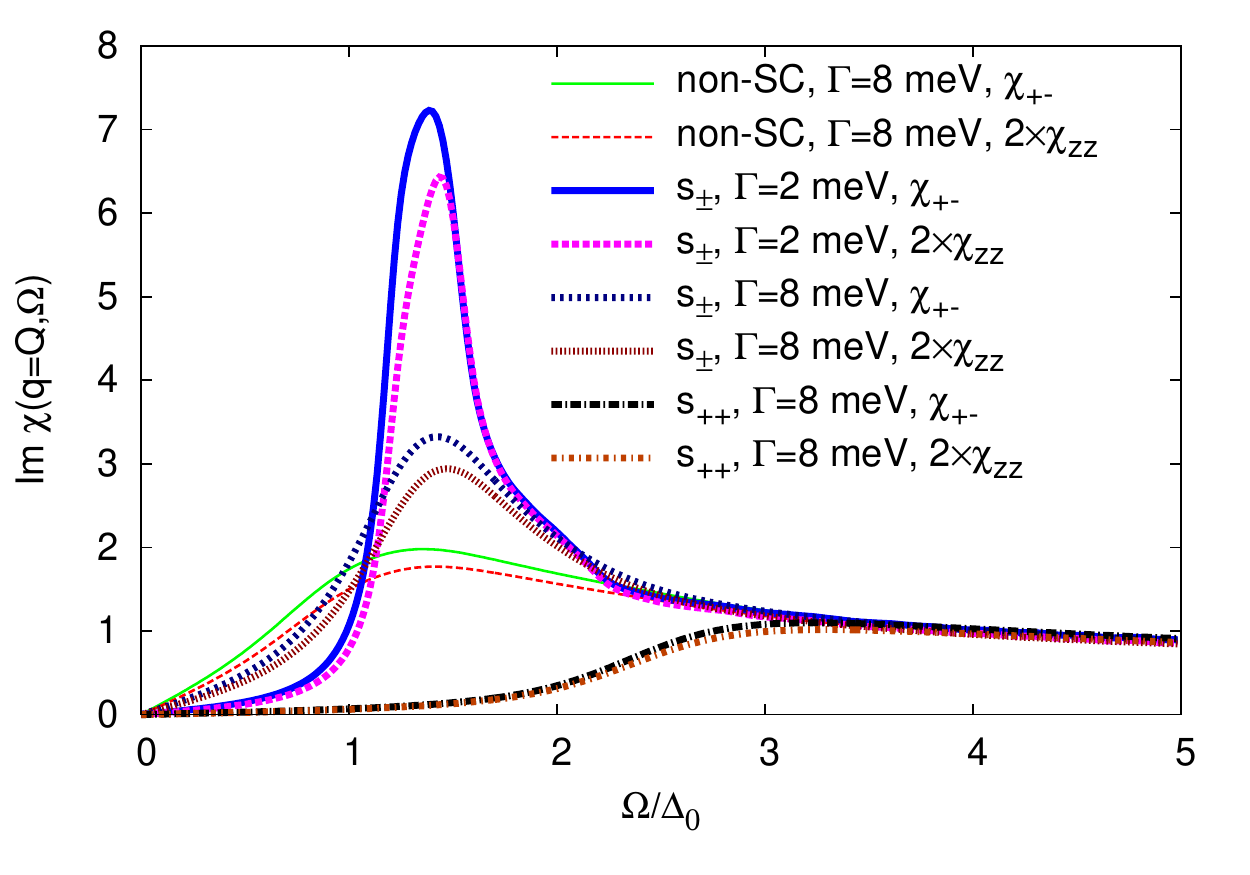}
\end{center}
\caption{Calculated $\mathrm{Im}\chi(\mathbf{Q},\Omega)$ with $\mathbf{Q}=(\pi,0)$ in the 1-Fe BZ in the normal state, and for the $s_{++}$ and $s_\pm$ pairing symmetries. In the latter case, the resonance is clearly seen below $\Omega = 2\Delta_0$.
\label{fig:ImChi3orbSO}}
\end{figure}

Now we switch to the three-orbital model. We calculated both $+-$ and $zz$ components of the spin susceptibility and confirmed that in the non-superconducting state $\chi_{+-} > 2\chi_{zz}$ at small frequencies, see Fig.~\ref{fig:ImChi3orbSO}. For the $s_\pm$ superconductor we observe a well defined spin resonance and $\chi_{+-}$ is again larger than $2\chi_{zz}$~\cite{Korshunov3orbSO}. Interestingly, for the $s_{++}$ state the disparity between $\chi_{+-}$ and $2\chi_{zz}$ is extremely small. With increasing impurity scattering rate the spin resonance peak broadens and its energy shifts to higher frequencies. This is similar to the results in the four-band model so we conclude that the orbital character and the SO coupling do not have much effect on the impurity-induced smearing of the spin resonance within the present approximation for the quasiparticle self-energy. Note that the spin response in the $s_\pm$ state is still distinct from the one in the $s_{++}$ state even for a sizeable value of $\Gamma$. This is important for the discussion of inelastic neutron data. Since all real materials are prone to disorder the natural question arise: is it possible to distinguish between $s_\pm$ and $s_{++}$ states in the presence of non-magnetic impurities looking at the neutron data? Here we demonstrate that the answer is yes, spin responses would be quite different. And the other important difference comes from the negligible disparity of $\chi_{+-}$ and $2\chi_{zz}$ components in the $s_{++}$ state, that contradicts results of the polarized neutron data~\cite{Lipscombe2010}.

\section{Conclusion}

We analysed the spin response in the superconducting state of FeBS in the presence of nonmagnetic disorder. The disorder was treated in the simple static Born approximation thus giving only basic qualitative trends. Average impurity scattering rate $\Gamma$ was considered as a parameter. For the small $\Gamma$, the spin resonance peak is clearly observed below the energy of $2\Delta_0$ and with increasing $\Gamma$ it becomes broader and almost vanishes once $\Gamma$ becomes comparable to $\Delta_0$. The energy of the spin resonance $\omega_s$ (determined as the maximum of the spin susceptibility) shifts to higher frequencies with increasing disorder. The spin resonance peak gains anisotropy in the spin space due to the spin-orbit coupling, so for the $s_\pm$ superconductor $\chi_{+-}$ is larger than $2\chi_{zz}$. On the other hand, for the $s_{++}$ state the disparity between transverse and longitudinal components is negligible. The spin response in the $s_\pm$ state is still distinct from that in the $s_{++}$ state even for a sizeable value of $\Gamma$.

\begin{acknowledgements}
We acknowledge partial support by the RFBR (grant 13-02-01395), President Grant for Government Support of the Leading Scientific Schools of the Russian Federation (NSh-2886.2014.2), and The Ministry of education and science of Russia (GF-2, SFU).
\end{acknowledgements}

\end{document}